# Emergence of insulating ferrimagnetism and perpendicular magnetic anisotropy in 3d-5d perovskite oxide composite films for insulator spintronics

Zeliang Ren[1,2,3†], Bin Lao[2,3†], Xuan Zheng[2,3,4], Lei Liao[5], Zengxing Lu[2,3], Sheng Li[2,3], Yongjie Yang[2,3], Bingshan Cao[1,2,3], Lijie Wen[2,3,6], Kenan Zhao[2,3], Lifen Wang[5], Xuedong Bai[5], Xianfeng Hao[6], Zhaoliang Liao[7*], Zhiming Wang[2,3,8*] and Run-Wei Li[2,3,8]

[1]Nano Science and Technology Institute, University of Science and Technology of China, Hefei 230026, Anhui, China

[2]CAS Key Laboratory of Magnetic Materials and Devices, Ningbo Institute of Materials Technology and Engineering, Chinese Academy of Sciences, Ningbo 315201, China

[3]Zhejiang Province Key Laboratory of Magnetic Materials and Application Technology, Ningbo Institute of Materials Technology and Engineering, Chinese Academy of Sciences, Ningbo 315201, China

[4]New Materials Institute, University of Nottingham Ningbo China, Ningbo 315100, China

[5]Beijing National Laboratory for Condensed Matter Physics, Institute of Physics, Chinese Academy of Sciences, Beijing 100190, China

[6]Key Laboratory of Applied Chemistry, College of Environmental and Chemical Engineering, Yanshan University, Qinhuangdao 066004, China

[7]National Synchrotron Radiation Laboratory, University of Science and Technology of China, Hefei 230026, Anhui, China

[8]Center of Materials Science and Optoelectronics Engineering, University of Chinese Academy of Sciences, Beijing 100049, China

**Abstract**: Magnetic insulators with strong perpendicular magnetic anisotropy (PMA) play a key role in exploring pure spin current phenomena and developing ultralow-dissipation spintronic devices, thereby it is highly desirable to develop new material platforms. Here we report epitaxial growth of $La_{2/3}Sr_{1/3}MnO_3$ (LSMO)-$SrIrO_3$ (SIO) composite oxide films (LSMIO) with different crystalline orientations fabricated by sequential two-target ablation process using pulsed laser deposition. The LSMIO films exhibit high crystalline quality with homogeneous mixture of LSMO and SIO at atomic level. Ferrimagnetic and insulating transport characteristics are observed, with the temperature-dependent electric resistivity well fitted by Mott variable-range-hopping model. Moreover, the LSMIO films show strong PMA. Through further constructing all perovskite oxide heterostructures of the ferrimagnetic insulator LSMIO and a strong spin-orbital coupled SIO layer, pronounced spin Hall magnetoresistance (SMR) and spin Hall-like anomalous Hall effect (SH-AHE) were observed. These results illustrate the potential application of the ferrimagnetic insulator LSMIO in developing all-oxide ultralow-dissipation spintronic devices.

# INTRODUCTION

In order to satisfy the ever-increasing demanding of information storage capacity and processing speed, reducing the energy consumption of electronic devices becomes more and more important in the microelectronics industry. In contrast to conventional microelectronics relying only the charge of electrons, spintronics use the additional spin of electrons to encode, store, process and transmit data, which is very promising to bring new capabilities to microelectronic devices[1-3]. Especially, spintronics based on ferromagnetic/ferrimagnetic magnetic insulators (FMI) has attracted tremendous attention for developing ultralow-dissipation devices. The local magnetic moments in FMI act as ideal media for pure spin angular momentum propagating while the thermal consumptions such as Joule heating can be efficiently avoided by suppressing the moving of electrons[4-6]. Additionally, magnets with perpendicular magnetic anisotropy (PMA), where the magnetic easy axis preferentially points toward normal to the surface, is essentially required to reduce the devices size while maintaining high thermal stability[7-11]. Combination of FMI and PMA further provides exciting functionalities for lowering current threshold and enhancing mobility of domain wall displacement, as well as achieving long decay length during spin-wave propagation[12]. Therefore, the development of FMI with PMA is much meaningful for the development of high-performance spintronic devices.

Magnetic insulators are frequently found in transition metal oxides with various crystal structures, such as rocksalt, perovskite, spinel, and garnet. Among them, perovskite $ABO_3$ oxides are particularly promising candidates to achieve high-quality all-oxides spintronic devices because of their versatile functionalities, including spin source generator, spin current detector and so on[13, 14]. However, magnetic insulators in perovskites often favor antiferromagnetism instead of ferromagnetic ordering, due to the super-exchange interaction between neighboring magnetic ions according to Goodenough-Kanamori-Anderson rules[15]. Moreover, FMI with strong PMA in perovskite oxides are rarely reported. Manganite $La_{1-x}Sr_xMnO_3$ (LSMO), as an archetypal colossal magnetoresistance materials, have been widely studied on account of its rich phase diagrams where multiple electric and magnetic properties are present[16, 17]. Previous studies had demonstrated that PMA can be obtained in LSMO films and heterostructures through engineering the lattice and orbital degrees of freedom[18-20]. Recently, strong PMA can be triggered in LSMO when interfacing with strong spin-orbital coupled $SrIrO_3$ (SIO) [21, 22]. However, the LSMO maintained metallic in most cases, unless large strain was imposed [23-25]. Through A- or/and B-site cation substitutions in $ABO_3$ the spin polarization of band structure can be directly controlled, leading to modified exchange interactions and associated magnetic and electrical ground states[26]. Thanks to the advances in epitaxial synthesis techniques, composite films have emerged as powerful platforms for realizing FMI and strong PMA in LSMO-based perovskite oxides since deliberating control of specific atoms can be realized[27]. Given these reasons, emergent FMI with strong PMA can be obtained via engineering cation substitution in LSMO and SIO, which enables to explore pure spin current phenomena in all-oxide heterostructures.

In this letter, we reported a facile method to fabricate LSMO-SIO (LSMIO) composite oxide films that are mixed homogenously at atomic level. The LSMIO films are FMI with PMA, whose temperature-dependent resistivity was fitted well by Mott variable-range-hopping model. By further constructing heterostructures of FMI LSMIO and strong spin-orbital coupled SIO layers, we explored pure spin current phenomena and observed pronounced spin Hall magnetoresistance (SMR) and spin Hall-like anomalous Hall effect (SH-AHE). These results demonstrate that FMI LSMIO is promising for exploring insulator spintronics in all oxides heterostructures and developing low-dissipation spintronic devices.

**RESULTS AND DISCUSSION**

Figure 1(a) shows the schematics of sequential two-target deposition process of LSMIO films using pulsed laser deposition (PLD). During the growth, the LSMO and SIO targets were rotated repeatedly and periodically. In each repeat, sub-monolayer 0.3 unit cell (u.c.) LSMO and 0.2 u.c. SIO were deposited, ensuring homogeneous mixture of two ceramic targets at the atomic scale. The LSMIO films have a nominal stoichiometry of $La_{0.4}Sr_{0.6}Mn_{0.6}Ir_{0.4}O_3$. The crystalline orientation of LSMIO films can be controlled by changing substrate orientation. The crystalline structures were characterized by X-Ray diffraction (XRD) $2\theta\text{-}\theta$ scans as shown in Fig. 1(b). The peaks of LSMIO films are labeled and are located on the left side of $SrTiO_3$ (STO) substrates closely. This result indicates that the films are slightly compressed (<0.1%) and of high-quality single crystal without impurity phases. To further characterize the interface quality and structural homogeneity, scanning transmission electron microscopy (STEM) measurements on the (001)-oriented LSMIO films, as shown in Figs. 1(c)-1(h), were performed. The sharp contrast across the LSMIO and STO substrates in Fig. 1(c), supplemented by a high-angle annular dark-field (HAADF) STEM image of the LSMIO films in Fig. 1(d), indicates high crystalline quality of the films. Figures 1(e)-(h) show the corresponding energy dispersive X-ray spectroscopy (EDS) mappings of La, Sr, Mn and Ir elements. All measured elements are distributed homogeneously at atomic level in the composite films without observable clustering regions.

To investigate the magnetic properties of LSMIO composite films, we measured magnetic hysteresis (*M-H*) loops and temperature-dependent magnetization (*M-T*) curves. Figures 2(a)-2(c) display the *M-H* loops observed with the magnetic field applied parallel ($H \perp c$) and perpendicular ($H // c$) to the film plane for differently oriented LSMIO composite films. The coercive field is found to be 0.67 T, which is much larger than that of soft ferromagnetic LSMO films. On the other hand, we note that the saturation moments are determined to be around 2.4 $\mu_B$/u.c., which are considerably smaller than that in LSMO films[28]. The reduction of the saturation moments can be explained by the antiferromagnetic coupling in the Mn-O-Ir bonds as evidenced from the first-principle calculations shown below, which leads to the ferrimagnetic ground state in LSMIO composite films. Figure 2(d) show the corresponding magnetization M as a function of temperature. The Curie temperature ($T_C$) becomes lower for all the LSMIO films as compared to pure LSMO bulk material. Intriguingly, by comparing the magnetic hysteresis loops measured under field parallel and perpendicular to films respectively, it is found that all easy magnetization axes are perpendicular to the film plane for all LSMIO films regardless of the crystal orientation while the magnetic anisotropy energy are $1.21 \times 10^6$, $1.55 \times 10^6$ and $1.56 \times 10^6$ erg/cm$^3$ respectively, which are comparable with the PMA reported in LSMO/SIO heterostructures[22]. Therefore, it is feasible to obtain perovskite FMI with strong PMA in LSMIO composite films. Considering the negligible compressive strain imposed by the SrTiO$_3$ substrates, the observed strong PMA is not due to magneto-elastic anisotropy as been observed previously in LSMO films or heterostructures when under large compressive strain[24, 25]. The strong PMA is due to the magnetic crystalline anisotropy originated from strong spin-orbital coupling and Mn-O-Ir bonding[21, 22, 29]

To further understand the electrical properties of LSMIO composite films, we have performed the temperature-dependent electric resistivity measurements. As shown in Fig. 2(e), the resistivity increases monotonically as the temperature decreases, indicating that all the LSMIO composite films are insulators. This is in sharp contrast to individual LSMO and SIO films which are metallic and semi-metallic in their bulk counterparts, respectively. To illustrate the origin of the emergent FMI state, we have fitted the resistivity data with thermal activation, Efros–Shklovskii variable-range-hopping (ES-VRH) and Mott variable-range-hopping (Mott-VRH) models, as shown in Figs. S3 (b)-(c) respectively. Figure 2(f) depicts the linear fitting curves using Mott-VRH model, which agrees well to the measured resistivity. Here, the Mott-VRH model is expressed as the following equation:

$$\rho(T) = \rho_{(0)} \, exp\left(T_M/T\right)^{(1/d+1)}$$

Where $\rho_0$ is resistance coefficient, $T_M$ is characteristic temperature and the exponent $d$ depends on the VRH mechanism. For the Mott VRH conduction, the exponent $d$ is dimension dependent and has a value of $d=3$ in a three-dimensional system[30-32]. From the fitted characteristic temperature $T_M$, we can infer the Mott hopping energy ($E_M$) as:

$$E_M = \frac{1}{4} k_B T \left(T_M/T\right)^{1/4}$$

The fitted hopping energy for (001)-, (110)- and (111)-oriented LSMIO composite films are determined to be 72.2, 93.7 and 117.2 meV at room temperature (300 K), respectively. The Mott-VRH model describes that the localized electron at the Fermi level moves to another localized state in an optimum hoping distance which is determined by the tradeoff between the lowest energy differences and the shortest hopping distances in a disordered system[33]. Given that the Mn-O-Ir bonds are dominated in the LSMIO composite oxide films, the Mott-VRH behavior suggests that the carriers hopping through the Mn-O-Ir bonds are suppressed and form localized states. This situation is in sharp contrast with metallic Mn-O-Mn bonds in LSMO films, where the electrons can itinerantly hop between neighboring $Mn^{3+}$ and $Mn^{4+}$ ions through the double-exchange mechanism[17, 34].

To gain further insight into the electronic and magnetic properties of LSMIO nanocomposite films, we have performed first-principles calculations based on the double perovskite model $LaSrMnIrO_6$ (Fig. 3(a)). From the total energy results of the distinct magnetic solutions (schematically shown in Fig. 4S and Tab. S2 in Supplementary Materials), we found that the ferrimagnetic, uncompensated antiferromagnetic alignment, due to the unequal magnetic moments on Mn and Ir sites, is the most stable state for LSMIO, affirming the experimental observations. Moreover, the GGA+U schemes prefer quite stable magnetic solution with magnetic moments of Mn (~ 4.3 μB) and Ir (~ -0.9 μB) ions, accompanied by considerable induced magnetic moments (~ -0.1 μB) for all the surrounding $O^{2-}$ anions as demonstrated in Fig. 3(b). Unexpected, the magnetic moment at Mn sites is slightly larger than the ideal spin-only 4 $\mu_B$ for $Mn^{3+}$ cations, suggesting the special magnetic coupling mechanism behind. Figure 3(c) shows that the ferrimagnetic ground state is insulating with a band gap of 0.2 eV, which agrees well with the experimental temperature-dependent electric resistivity data. The antiferromagnetic coupling and unexpected large magnetic moment on $Mn^{3+}$ site can be interpreted in the framework of superexchange interactions through the virtual hopping bridged by O $2p$ spin-up electrons, as illustrated in Fig. 3(d), which is derived from the partial density of states and spin density plots. We can see that the virtual hopping from Ir $t_{2g}^{\uparrow}$ to the empty Mn $e_g^{\uparrow}$ via the bridged O $2p^{\uparrow}$ states accounts for the antiferromagnetic coupling and the abnormal large computed $Mn^{3+}$ magnetic moment as well as the exclusively negative sign of the induced $O^{2-}$ magnetic moments.

Finally, to explore the application of FMI LSMIO in pure spin current phenomenon, we fabricated a LSMIO/SIO heterostructure and performed temperature-dependent magnetoresistance measurements. Figures 4(a)-4(b) show schematic structure of Hall bar devices and measurement geometry. As illustrated in Fig. 4(a), the heterostructure consists of the FMI LSMIO and a strong spin-orbital coupled SIO layer. The SIO layer can efficiently convert the charge current into spin current[35, 36], so a pronounced SMR can be expected due to the asymmetry between absorption and reflection of the spin current at the LSMIO/SIO interface. Figure 4(c) depicts the angle dependent magnetoresistance measurements where external field rotates in the y-z plane around angle β. It is necessary to distinguish the difference between SMR and anisotropic magnetoresistance (AMR). SMR signal is known to be described as:

$$\rho_{xx} = \rho_0 - \Delta\rho m_y^2$$

while AMR signal is described as:

$$\rho_{xx} = \rho_0 - \Delta\rho m_x^2$$

In these two formulas, the $\rho_0$ is constant resistivity offset, $\Delta\rho$ is amplitude of the resistivity change as a function of the magnetization orientation and $m_i$ (i = x, y) is the projections of the magnetization orientation unit vector along x and y axis in the coordinate system[37-39]. Note that $\rho_{xx} = v_{xx}wt/J_q l$ and $\rho_{xy} = v_{xy}t/J_q$, where $w$ is channel width ($w$ = 10 um), the $l$ is the length of the separated voltage contacts ($l$ = 50 um) and the $t$ is the thickness of SIO layer ($t$ = 5 nm). During the measurement, the external field was fixed to 4 T which is much larger than the coercive field of LSMIO (0.67 T) in order to obtain collinear magnetization. Figure 4(c) shows that the resistivity data exhibit the $\cos^2\beta$ dependence. This is in consistent with the SMR signal since $m_y$ has cosβ dependence, while AMR keep constant in angle β dependent measurements since $m_x$ equals to 0. These measurements demonstrate that the SMR can be detected in all perovskite oxide heterostructures.

Figure 4(d) shows the Hall resistance measurements with the magnetic field applied perpendicular to the sample and varied from -4 T to 4 T. By subtracting a linear contribution from ordinary Hall effect, a clear hysteresis loop curve is visible, which resembles the AHE. Figure 4(e) shows the temperature-dependent Hall resistance after subtracting a linear term. The coercive field inferred from the Hall hysteresis loop is about 0.67 T at 10 K, which is similar as that of LSMIO films measured by SQUID shown in Fig. 2(a). Moreover, the coercive field decreases as a function of temperature and becomes invisible at above 120 K. These results demonstrate that the evolution of AHE with temperature resembles that of the magnetization versus temperature curve as shown in Fig. 2(d), indicating that the observed AHE is closely related to the PMA of LSMIO composite films. We note that interfacial magnetism and AHE has been intensively studied in LSMO/SIO superlattices and heterostructures[21, 40-46]. Our observed AHE is quite different from that the AHE reported in ref. 41, where a large AHE is originated from SIO layers due to magnetic proximity effect. Our AHE stems from the reflection of the spin current at the interface where an out-of-plane component ($m_z$) of the LSMIO magnetization rotates the spin orientation of the spin current and generates a transverse voltage via the inverse spin Hall effect (ISHE). Such AHE is termed as SH-AHE[47, 48]. Therefore, both SMR and SH-AHE indicate occurrence of the spin transfer at the interface of the conductor and ferrimagnetic insulator. The spin current can be injected into insulators which gives a possibility of transmitting spin information through an insulator. Such behavior would have an advantage of ultralow power dissipation due to absence of Joule heating.

**CONCLUSION**

In conclusion, we have synthesized LSMIO composite oxide films with different crystalline orientations by sequential two-target deposition process. The LSMIO composite films exhibit high crystalline quality with homogeneous mixture of LSMO and SIO at atomic level. The electric and magnetic measurements, complemented by first-principle calculations, indicate that the LSMIO composite films are FMI with strong PMA. The temperature-dependent electric resistivity is well described by Mott-VRH model. Furthermore, through fabricating heterostructures of the FMI LSMIO and a strong spin-orbital coupled SIO layer, we observed pronounced SMR and SH-AHE. These results illustrate the potential application of LSMIO in developing all-oxide ultralow-dissipation spintronic devices. Given the scarcity of perovskite FMI with strong PMA, our result provides a promising real material candidate to boost the development of insulating spintronics devices.

**EXPRIMENTAL SECTION**

*Sample preparation:* LSMIO composite films were synthesized by PLD, equipped with KrF ($\lambda$=248 nm) excimer laser and high reflection high energy electron diffraction (RHEED). Prior to the film deposition, (001)-, (110)- and (111)-oriented STO substrates were etched by $NH_4F$-buffered hydrofluoric acid followed by annealing at 1100 °C for 90 minutes. During films deposition, the temperature of substrates, oxygen partial pressure and laser fluence were set to 700 °C, 0.1 mbar and 1.5 J/cm$^2$ at the frequency of 2 Hz, respectively. The film growth was monitored by RHEED and the thickness was determined to be 12 nm.

*Sample characterizations:* The quality and structure of the films was characterized by XRD (Bruker), STEM and EDS (JEOL Grand ARM 300), respectively. The magnetic and electric properties were measured by Quantum Design Magnetic Property Measurement System (MPMS-SQUID, Quantum Design) and Physical Property Measurement System (PPMS, Quantum Design). Further investigations of spin-related SMR and AHE were performed on Hall bar devices composed of LSMIO and SIO using a home-build low-temperature and high-intensity magnetic field magneto-electric system. The LSMIO/SIO heterostructures were prepared by PLD as the same condition with LSMIO films. The as grown (001)-oriented LSMIO/SIO heterostructure was patterned into Hall bars by photolithography and Ar ion etching.

*First-principle calculations:* In order to describe the structure of LSMIO composite films, we constructed the canonical perovskite-type model within 2×2×2 supercell ($La_4Sr_4Mn_4Ir_4O_{24}$, 40 atoms). The spin-polarized density functional theory (DFT) calculations are achieved with the plane wave basis set as implemented in the Vienna *ab initio* simulation package[49, 50]. For the exchange correlation functional, the generalized gradient approximation (GGA) is used according to the Perdew-Burke-Ernzerhof scheme[51], and GGA+U approaches[52, 53] are also performed for including the static electron correlation, with the Hubbard U = 4 eV (2 eV) and Hund J = 0.8 eV (0.4 eV) for Mn 3*d* (Ir 5*d*) states. A plane-wave energy cutoff of 400 eV was employed. The *k*-point meshes over the total Brillouin zone were sampled by 4×4×4 and 6×6×6 grids constructed according to the Monkhorst-Pack scheme[54, 55] for structural relaxations and electronic calculations, respectively. We optimized the lattice vectors, volumes, as well as internal atomic with the relaxation terminated once the total energies and residual atomic forces converged to < 1×10$^{-5}$ eV and < |0.01| eV/Å, respectively.

■ **AUTHOR INFORMATION**

**Corresponding Author**

zliao@ustc.edu.cn.

zhiming.wang@nimte.ac.cn.

**Author contribution**

■ **ACKNOWLEDGMENTS**

This work was supported by the National Key Research and Development Program of China (Nos. 2017YFA0303600, 2019YFA0307800), the National Natural Science Foundation of China (Nos. 12174406, U1832102, 11874367, 51931011, 51902322), the Key Research Program of Frontier Sciences, Chinese Academy of Sciences (No. ZDBS-LY-SLH008), the Thousand Young Talents Program of China, K.C.Wong Education Foundation (GJTD-2020-11), the 3315 Program of Ningbo, the Natural Science Foundation of Zhejiang province of China (No. LR20A040001), the Beijing National Laboratory for Condensed Matter Physics.

■ **REFERENCES**

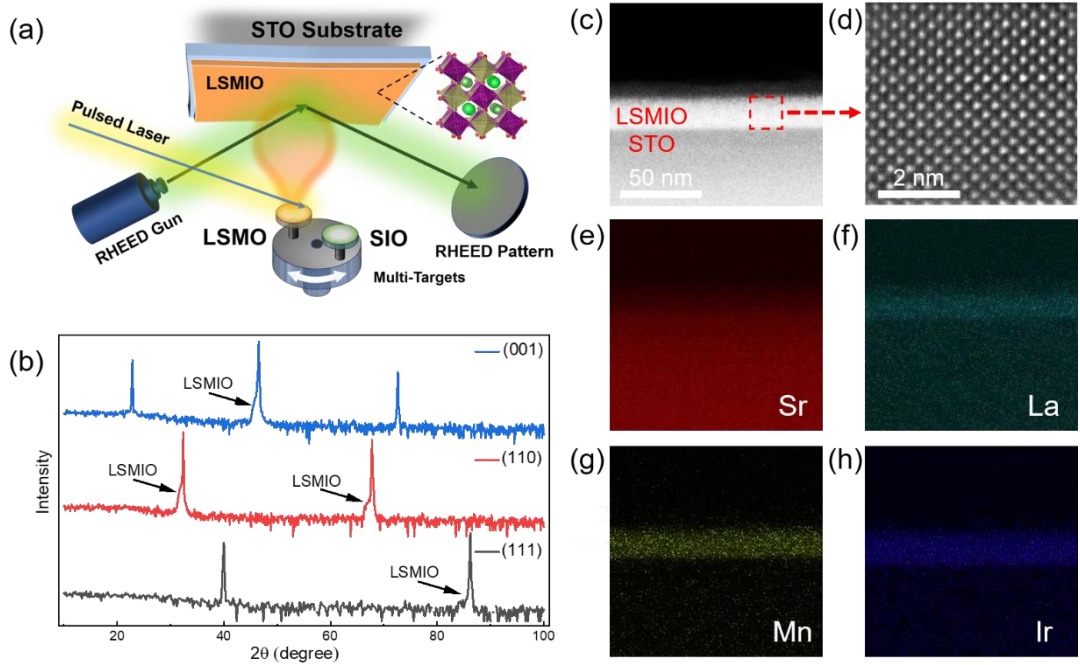

Figure 1. (a) Schematic illustration of sequential two-target deposition process of LSMIO films. (b) XRD 2θ-θ scans of LSMIO films with (001), (110) and (111) crystalline orientations. (c) HAADF-STEM images of the (001)-oriented LSMIO/STO. The high-magnification image in (b) correspond to the single crystalline LSMIO films. (e)-(h) The corresponding EDS mappings of La, Mn, Ir and Sr elements, respectively, confirming cation distribution homogeneously at atomic level.

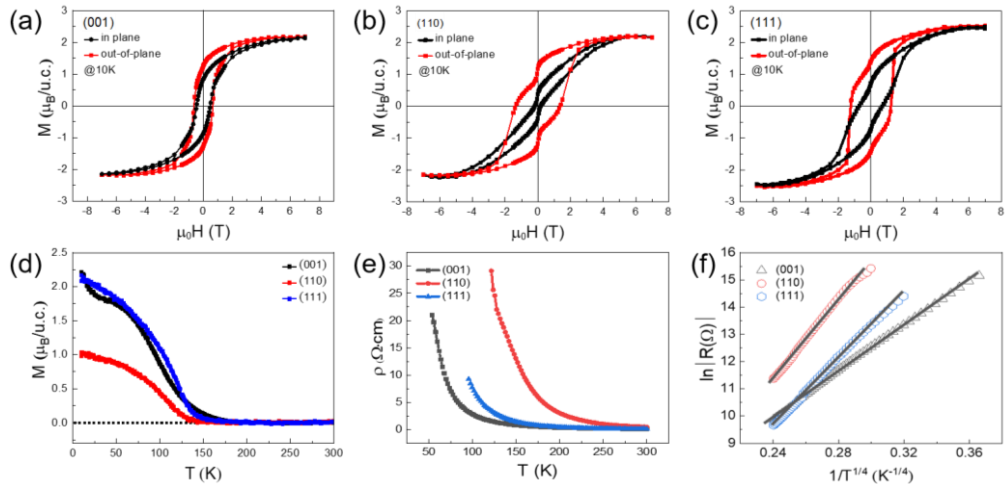

Figure 2. (a)-(c) Magnetic hysteresis loops measured with the magnetic field H applied along in-plane (black) and out-of-plane (red) for LSMIO composite films with different crystalline orientations. (d)-(e) The corresponding magnetization M and electric resistivity as a function of temperature T for different oriented LSMIO composite films. (f) The resistivity data are plotted with $\ln R \sim 1/T^{1/(d+1)}$ ($d = 3$) to fit with the three-dimensional Mott-VRH model.

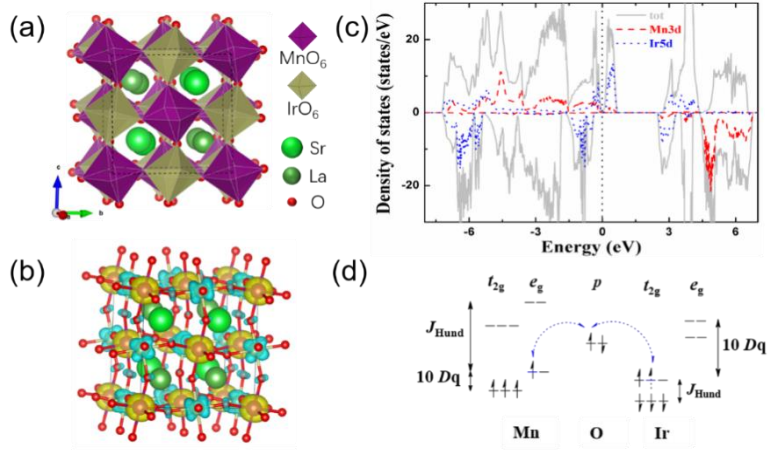

Figure 3. (a) Double perovskite structure of LaSrMnIrO$_6$, in which the Mn atoms (medium purple spheres) and the Ir atoms (medium brown ones) are surrounded by O$_6$ octahedron denoted by small red spheres, while the light (dark) green spheres are Sr (La) atoms. (b) Isosurfaces of the spin density (0.02 e/Å$^2$) for double double perovskite LaSrMnIrO$_6$, within the ferrimagnetic magnetic ground state. Yellow and cyan color denote spin up and spin down, respectively. (c) Total and partial density of states of Mn-3d and Ir-5d orbitals for LaSrMnIrO$_6$. Majority and minority spin are presented in the top and bottom channels, respectively. (d) Schematic representation of the superexchange interaction through the Mn-O-Ir bonds. The virtual electron hopping results into the antiferromagnetic coupling between Mn-O-Ir bonds.

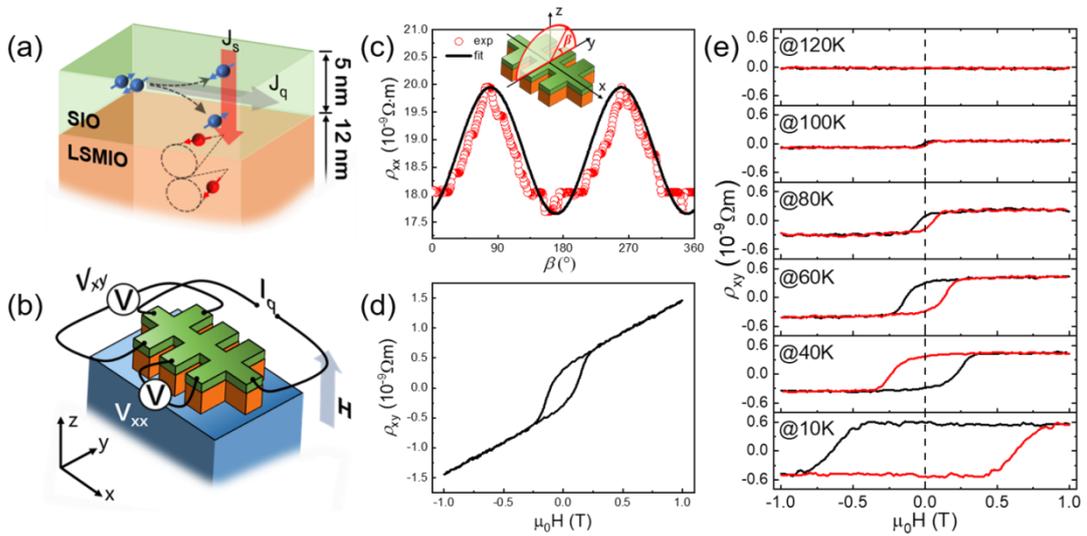

Figure 4. (a) Illustration of the SMR in LSMIO/SIO heterostructure. (b) Schematic sample structure and measurement geometry. (c) The angle dependent SMR signal. The inset shows the measurement geometry with external field rotating in the y-z plane around angle β. (d) Transverse Hall resistivity $\rho_{xy}$ measured with the magnetic field applied perpendicular to the film plane; (e) Temperature dependent AHE sign

**Supplemental Information**

1. Structural characterization of LSMIO and LSMIO/SIO heterostructure.

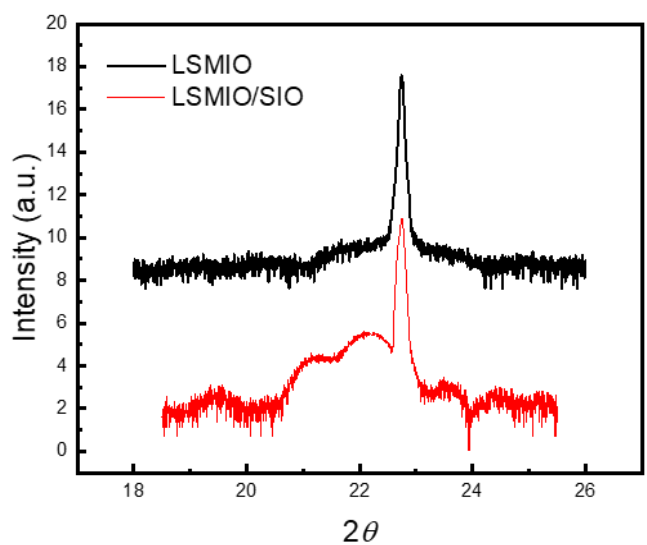

**Figure S1.** XRD scans of LSMIO and LSMIO/SIO heterostructure.

As shown in Fig. S1, the peak of SIO and LSMIO can be seen clearly in LSMIO/SIO heterostructure while there is only one peak of LMSIO in single LSMIO layer which means there is only one single crystal of LSMO-SIO composite films.

2. Structural characterization of LSMIO composite films by STEM

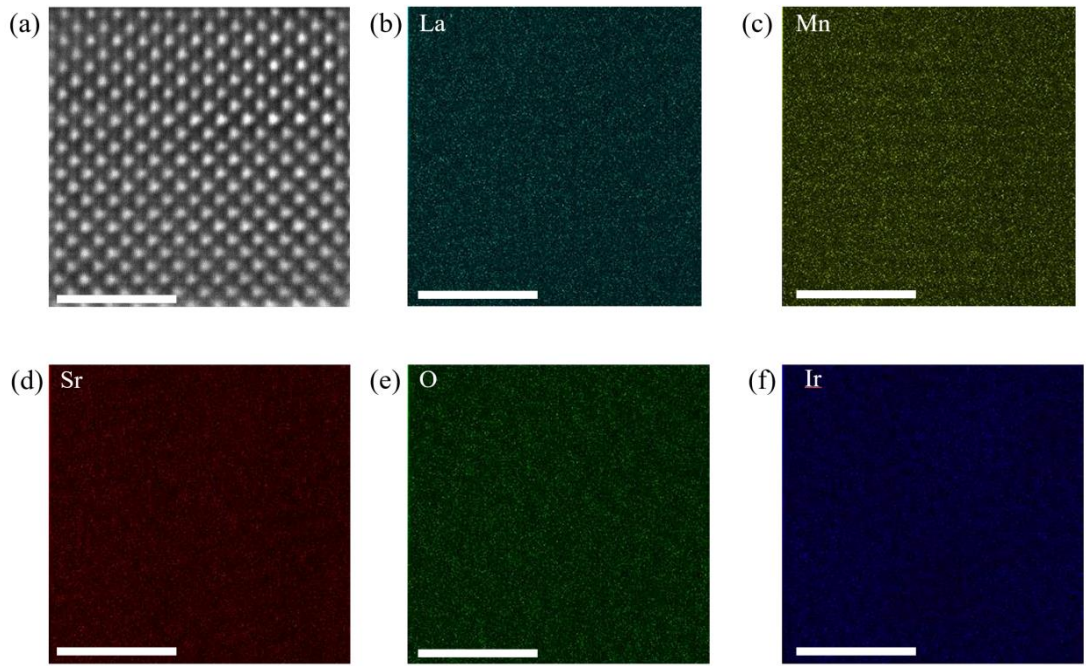

**Figure S2** (a) High-resolution HAADF-STEM image of the LSMIO films (b-f) EDS maps of La, Mn, Sr, O and Ir for LSMIO films, Scale bar:2nm.

We can see the homogeneously mixture of atoms in LSMIO films in Figs S2(b)-(f).

3. The fittings of measured resistivity of LSMIO.

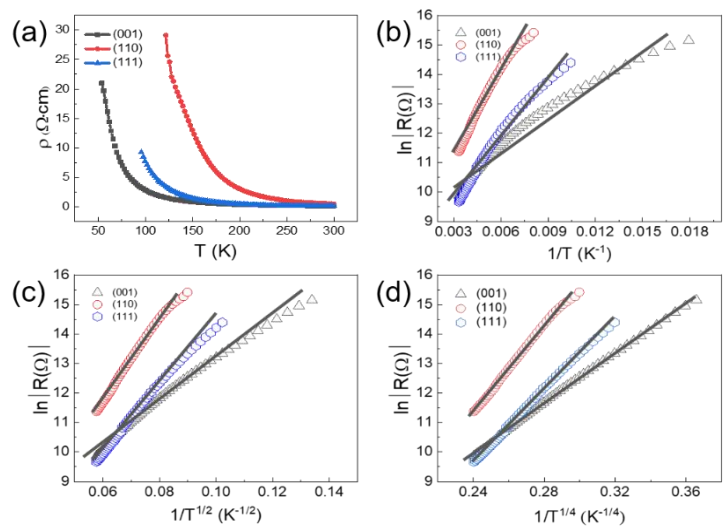

**Figure S3.** (a) Temperature-dependent electric resistivity of LSMIO composite films with different crystalline orientations. (b)-(d) The resistivity data are plotted with $\ln R \sim 1/T^{[1/(d+1)]}$ (d = 0,1,3) to fitted with the thermal activation, Efros–Shklovskii-VRH and Mott-VRH model respectively.

In this picture, we plot the resistivity with different physical model with thermal activation (d=0), Efros–Shklovskii-VRH (d=1) and 3-dimentional Mott-VRH (d=3). However, the nonlinearity of the curves in the Fig. S3(b)-(c) indicate that the thermal activation model cannot well fit the resistivities of LSMIO while only the Mott-VRH fits well to the measured resistivity which can be seen in Fig. S3(d).

4. DFT calculations.

In order to describe the structure of LSMIO composite films, we constructed the canonical perovskite-type model within 2×2×2 supercell ($La_4Sr_4Mn_4Ir_4O_{24}$, 40 atoms), with aid of the experimental characterizations, which can be pictured as a framework consisting of alternative corners sharing $MnO_6$ and $IrO_6$ octahedra separated by $La^{3+}/Sr^{2+}$ ions, due to the quite different charge states and ionic sizes of $3d$ and $5d$ cations. Concerning the $La^{3+}/Sr^{2+}$ ions distributions, the columnar $La^{3+}/Sr^{2+}$ cation order is taken (Fig. 3(a)), as it is prevailing among the nine possible double double AA′BB′O$_6$ perovskite[1]. Owning to the rotation and tilting of $MnO_6$ and $IrO_6$ octahedra, the Mn-O-Ir bond angles significantly deviate from 180º.

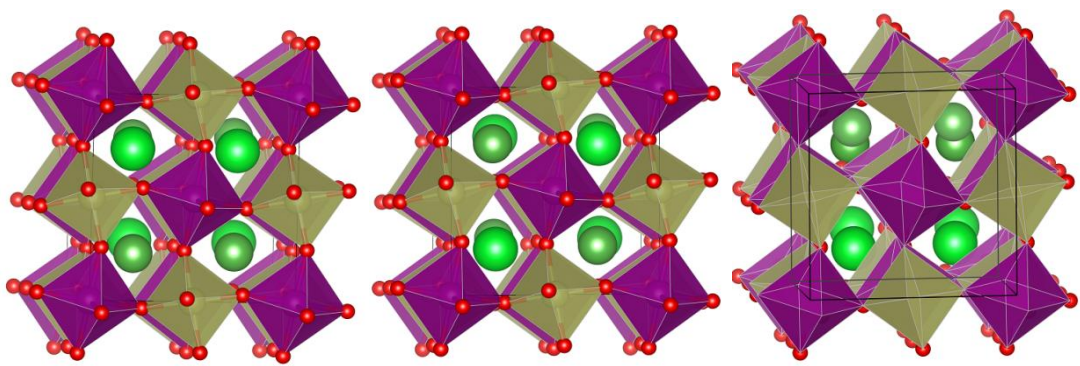

**Figure S4.** Different A-cations configurations of LSMIO composite considered in this study. (left) columnar type (middle) rocksalt type (right) layer type.

**Table S1.** Relative energies (meV/f.u.) obtained by GGA scheme within the magnetic ground state, i.e. ferrimagnetic ordering. The results suggest that the columnar type A-cations pattern is preferable in LSMIO composite films.

|  | Columnar | Rocksalt | Layer |
|---|---|---|---|
| Energy (meV) | 0.0 | 15.7 | 86.7 |

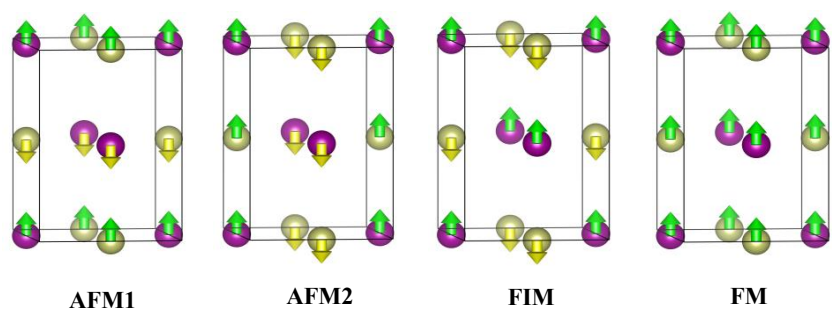

    AFM1        AFM2        FIM        FM

**Figure S5.** Different magnetic configurations of LSMIO composite considered in this study. Only Mn atoms (purple spheres) and the Ir atoms (brown ones) are shown for simplify. Green (yellow) arrows indicate spin up (down) character.

**Table S2.** The calculated total energies (eV/2×2×2 supercell) obtained by GGA and GGA+U ($U_{Mn}$ = 4.0 eV, $U_{Ir}$ = 2.0 eV) methods. The FIM (ferrimagnetic) is the ground state.

|  |  | FM | FIM | AFM1 | AFM2 |
|---|---|---|---|---|---|
| Energy (eV) | GGA | -302.0302 | -302.0391 | -301.9835 | -301.9835 |
|  | GGA+U | -286.4180 | -286.9023 | -286.2655 | -286.5611 |

We performed electronic calculations on four distinct magnetic alignments: ferromagnetic (FM), ferrimagnetic (FIM) (i.e., with Mn ions being antiferromagnetic (AFM) coupled to six neighboring Ir ions and vice versa) and two distinct AFM ones, within the normal GGA/GGA+U ($U_{Mn}$ = 4.0 eV, $U_{Ir}$ = 2.0 eV) prescriptions with the goal of identifying the magnetic ground state. From the total energy results of the distinct magnetic solutions, we found that the ferrimagnetic, uncompensated antiferromagnetic alignment, due to the unequal magnetic moments on Mn and Ir sites, is the most stable state for LSMIO, irrespective of whether the Hubbard U values was taken into account or not, affirming the experimental magnetic observations.

The conspicuous feature of electronic structure is that most of majority channel of Mn 3$d$ orbitals are occupied, forming a broad valence band ranged from -7.0 eV to the Fermi level, while the corresponding minority channel is entirely empty centered at 4.5 eV above the Fermi level, revealing that electronic configuration is close to the $t_{2g}^{3\uparrow}e_g^{1\uparrow}$ state and the oxidation state is 3+. Unexpected, the magnetic moment at Mn sites is slightly larger than the ideal spin-only 4 $\mu_B$ for $Mn^{3+}$ cations, suggesting the special magnetic coupling mechanism behind. For the Ir states, on one hand, the occupied minority Os $t_{2g}$ states are completely separated from the empty $e_g$ states by an energy interval of ~ 4 eV, accompanied by a moderate spin splitting of ~ 2 eV. On the other hand, the spin up $t_{2g}$ states are partially filled, splitting into a small band gap under the Hubbard U correlation effect. As a consequence, electronic configuration is close to the nominal $t_{2g}^{3\downarrow1\uparrow}$ state and the oxidation state is 5+. In addition, there are some Ir $e_g$ obitals present around -6 eV below the Fermi level drowned in the O 2$p$ manifold, unfolding the strong covalency between the Ir 5$d$ and O 2$p$ states. The covalent bonding is so strong that the computed Ir magnetic moment (~ 0.9 $\mu_B$) is remarkable smaller than the ideal spin-only value (~2 $\mu_B$).

The antiferromagnetic coupling and unexpected large magnetic moment on $Mn^{3+}$ site can be interpreted in the framework of superexchange interactions through the virtual hopping bridged by O 2$p$ spin-up electrons. We can see that in contrast to the large energy separation of more than 5 eV between the down-spin $Mn^{3+}$ and $Ir^{5+}$ $t_{2g}$ states, the up-spin $Mn^{3+}$ $e_g$ and $Ir^{5+}$ $t_{2g}$ states are separated by on the small band gap of 0.2 eV, and their hybridization via O 2$p$ up-spin orbitals can push the occupied $Mn^{3+}$ $t_{2g}^{3\uparrow}e_g^{1\uparrow}$ and $Ir^{5+}$ $t_{2g}^{3\downarrow1\uparrow}$ bands downwards, consequently stabilizes the antiferromagnetic interaction and then the ferrimagnetic ground state. In other words, the virtual hopping from $Ir^{5+}$ $t_{2g}^{\uparrow}$ to the empty $Mn^{3+}$ $e_g^{\uparrow}$ via the bridged O $2p^{\uparrow}$ states accounts for the antiferromagnetic coupling and the abnormal large computed $Mn^{3+}$ magnetic moment as well as the exclusively negative sign of the induced $O^{2-}$ magnetic moments.